\documentclass[epj]{svjour}
\usepackage{graphics}
\usepackage{graphicx}
\usepackage{amsmath}
\usepackage{rotating}
\usepackage{lipsum}
\usepackage{tikz}
\usepackage{hvfloat}

\usepackage{afterpage}

\newcommand{\pol}[1]{\mathaccent"017E{#1}}

\definecolor{referee1}{rgb}{0.0,0.0,0.0}

\begin{document}

\title{A comprehensive study of analyzing powers in the proton-deuteron break-up channel at 135~MeV\thanks{Supplementary material in the form of a pdf file available from the journal web page at http://*****}}

\author
{
M.~T.~Bayat\inst{1}\thanks{m.t.bayat@rug.nl}
\and 
H.~Tavakoli-Zaniani\inst{1,2}
\and
H.~R.~Amir-Ahmadi\inst{1}
\and
A.~Deltuva\inst{3}
\and
M.~Eslami-Kalantari\inst{2}
\and
J.~Golak\inst{4}
\and
N.~Kalantar-Nayestanaki\inst{1} 
\and 
St.~Kistryn\inst{5} 
\and 
A.~Kozela\inst{6} 
\and
H.~Mardanpour\inst{1}
\and
J.~G.~Messchendorp\inst{1}\thanks{j.g.messchendorp@rug.nl}
\and 
M.~Mohammadi-Dadkan\inst{1,7} 
\and 
A.~Ramazani-Moghaddam-Arani\inst{8}
\and
R.~Ramazani-Sharifabadi\inst{1,9} 
\and 
R.~Skibi\'nski\inst{4}
\and
E.~Stephan\inst{10} 
\and
H.~Wita\l{}a\inst{4}
}

\institute
{
KVI-CART, University of Groningen, Groningen, The Netherlands 
\and
Department of Physics, School of Science, Yazd University, Yazd, Iran 
\and
Institute of Theoretical Physics and Astronomy, Vilnius University, Saul\.etekio al. 3, 10222 Vilnius, Lithuania 
\and
M.~Smoluchowski Institute of Physics, Jagiellonian University, Krak\'ow, Poland 
\and
Institute of Physics, Jagiellonian University, Krak\'ow, Poland 
\and
Institute of Nuclear Physics, PAS, Krak\'ow, Poland 
\and
Department of Physics, University of Sistan and Baluchestan, Zahedan, Iran 
\and
Department of Physics, Faculty of Science, University of Kashan, Kashan, Iran 
\and
Department of Physics, University of Tehran, Tehran, Iran 
\and
Institute of Physics, University of Silesia, Chorz\'ow, Poland 
}

\date{Received: date / Revised version: date}
\abstract{
A measurement of the analyzing powers for the $^2$H$(\pol{p},pp)n$ break-up reaction was carried out at KVI exploiting a polarized-proton beam at an energy of $135$~MeV. The scattering angles and energies of the final-state protons were measured using the Big Instrument for Nuclear-polarization Analysis (BINA) with a nearly $4\pi$ geometrical acceptance. In this work, we analyzed a large number of kinematical geometries including forward-forward configurations in which both the final-state particles scatter to small polar angles and backward-forward configurations in which one of the final-state particles scatters to large polar angles. The results are compared with Faddeev calculations based on modern nucleon-nucleon (NN) and three-nucleon (3N) potentials. Discrepancies between polarization data and theoretical predictions are observed for configurations corresponding to small relative azimuthal angles between the two final-state protons. These configurations show a large sensitivity to 3N force effects.
} 

\titlerunning{A comprehensive study of analyzing powers in the proton-deuteron break-up channel at $135$~MeV} 
\authorrunning{M.~T.~Bayat et al.} 

\maketitle
\section{Introduction}
\label{intro}
Today's nucleon-nucleon (NN) potentials such as Argonne-V18 (AV18)~\cite{Wiringa95}, Reid-93~\cite{Stoks94}, Nijmegen-I and II~\cite{Stoks94} and CD-Bonn (CDB)~\cite{Machleidt96,Machleidt01} provide an excellent description of NN scattering observables and of the properties of the deuteron. However, exact calculations using two-nucleon forces (2NFs) alone are not sufficient to describe, with similar accuracy, systems consisting of more than two nucleons. For example, none of the NN potentials can reproduce the binding energy of the simplest three-nucleon system, the triton~\cite{Nogga2000}. A similar underbinding occurs for other light nuclei as well~\cite{Carlson15}. The most promising and widely-investigated solution is the addition of a three-nucleon force (3NF), a contribution that cannot be reduced to pair-wise reactions. The 3NFs arise in the framework of meson exchange theory where a 3N interaction can be derived by means of two-pion exchange between all three nucleons with an intermediate excitation of one of them to a $\Delta$-isobar such as in Urbana-IX (UIX)~\cite{Carlson83,Pudliner95} and Tucson-Melbourne (TM99)~\cite{Coon79,Coon01} models or they appear fully naturally in Chiral Perturbation Theory (ChPT) at a certain order of chiral expansion~\cite{Weinberg90,Weinberg91,Epelbaum06}. Alternatively, 3NFs can be included in a coupled-channel approach with an explicit $\Delta$-isobar excitation like the CDB+$\Delta$ (NN+3NF)~\cite{Deltuva2003a,Deltuva2003b}.

The importance of 3NF contributions to the dynamics of systems composed of more than two nucleons was first established in binding energies of few-nucleon states~\cite{Carlson15}. Further verification of the role of the 3NF has been carried out on the basis of scattering experiments. Various observables were measured in elastic nucleon-deuteron scattering and in the break-up of the deuteron via its collision with a nucleon. An extensive discussion of the present status of our understanding of the dynamics of the three-nucleon system, based on modern calculations and many precise and rich data sets, can be found in review articles~\cite{Sagara10,Nasser12,Kistryn13}. The 3NF turned out to be very important for improving the description of the cross section for nucleon-deuteron elastic scattering data. At beam energies above $100$~MeV per nucleon certain discrepancies between data and calculations still persist, though significantly reduced as compared to predictions based on purely NN potentials. The experimental data demonstrate both the successes and the difficulties of the current nuclear force models in describing cross sections, analyzing powers, spin-transfer and spin-correlation coefficients for N$d$ elastic scattering~\cite{Ela17}.

In the past three decades, many measurements have been carried out at KVI and at other laboratories to obtain high-precision data sets to provide a better understanding of the underlying dynamics of the 3NF. The experimental studies of the $^2$H$(\pol{p},pp)n$ reaction at $135$ and $190$~MeV~\cite{Mardanpour10,Eslami09} show a large (and growing with beam energy) discrepancy between the measured data and theoretical predictions for the vector analyzing power for a number of configurations. These discrepancies demonstrate that spin-dependent parts of the 3NFs are not completely understood~\cite{Ermisch05}. Based on these observations, and considering the rich phase space of the break-up reaction, it was decided to expand the analysis of the data taken in 2006 at KVI. In this work, we extended the earlier analysis~\cite{Eslami09,Eslami_thesis} that was done for kinematical configurations in which protons scatter to small forward angles up to $35^{\circ}$ by analyzing configurations at which one of the final-state protons scatters to the backward angles starting from $40^{\circ}$.
\begin{figure}[!h]
\centering
\includegraphics[angle=0,width=0.49\textwidth]{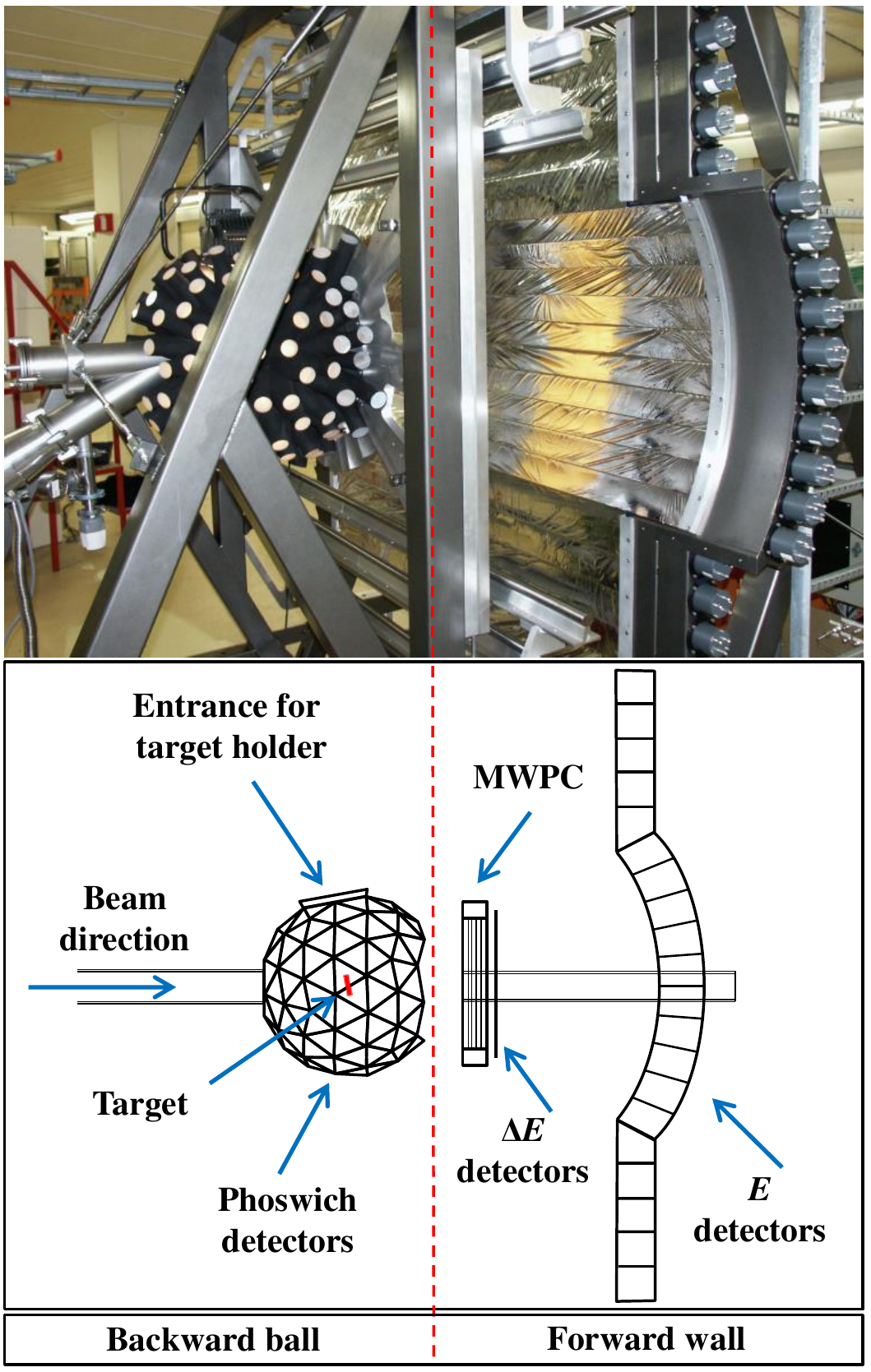}
\caption{A side view of BINA. The top panel shows a photograph of BINA's 
side-view and the bottom one presents schematic drawing of the forward
 wall and the backward ball.}
\label{BINA}
\end{figure}
\section{Experimental setup}
\label{Exp}
The experiment was performed at the Kernfysisch Versneller Instituut\footnote{Presently known as KVI-Center for Advanced Radiation Technology (KVI-CART).} (KVI) in Groningen, the Netherlands. A polarized proton beam produced by POL\-arized Ion Source (POLIS)~\cite{kremers97} was accelerated with the superconducting cyclotron AGOR (Acc\'el\'erateur Groningen ORsay)~\cite{Gales87} to $135$~MeV. The beam polarization was measured using a Lamb-shift polarimeter (LSP) in the low-energy beam line and by an in-beam polarimeter (IBP) that was installed at the high-energy beam line after acceleration~\cite{ahm1}. The proton beam impinged on a liquid-deuterium target and the reaction products were detected by the Big Instrument for Nuclear-polarization Analysis (BINA). The BINA detection system enables us to study break-up and elastic reactions at intermediate energies in almost 80\% of the full $4\pi$ solid angle coverage; see Fig.~\ref{BINA}. BINA is composed of two main parts, the forward wall and the backward ball. In the following, these two parts are briefly described.

The forward wall consists of three parts: a Multi-Wire Proportional Chamber (MWPC), $\Delta E$- and $E$-scintillators. The forward wall covers the polar angle ($\theta$) in the range of $10^{\circ}$-$32^{\circ}$ with full azimuthal-angle ($\phi$) coverage while, due to the corners of the MWPC, the azimuthal-angle coverage is limited for the polar angles from $32^{\circ}$ to $37^{\circ}$. When a particle passes through the MWPC, its coordinates are recorded. Subsequently, a small fraction of its energy is deposited in the $\Delta E$-scintillators. At the end of its trajectory, the particle stops inside of the $E$-scintillators if its energy is less than $140$~MeV (in case of protons). The type of particle can be identified by combining the information obtained from the $E$- and $\Delta E$ scintillators. All parts of the forward wall have a central hole for leading the beam pipe through the system. In the following subsections, these parts are described in more detail; see also Refs.~\cite{Mohammad_thesis,Mohammad_2019,Mohammad_FB22_2020}.

BINA's MWPC, with an active area of 38$\times$38~cm$^{2}$, is installed at a distance of 29.5~cm from the target position and it consists of 3 planes. For further details of BINA's MWPC, we refer to Ref.~\cite{Volkerts99}.

$E$-scintillators form the cylindrically-shaped part whose center coincides with the center of the target and two flat wing-like parts placed above and below the cylindrical part. The latter, which was not used in the present experiment, can be used for detecting the secondary scattered particles in polarization-transfer experiments. The cylindrical part consists of 10 horizontal scintillator bars with a trapezoidal cross section and the dimensions of $(9-10)\times 12\times 220$~cm$^{3}$ each. The two central scintillators have a hole in the middle for passage of the beam pipe. 

$\Delta E$-scintillators in combination with $E$-scintillators are used to identify the particle type (i.e. proton, deuteron etc.) as well as to determine the MWPC efficiency. The array of $\Delta E$-scintillators is composed of 12 thin slabs ($0.2\times 3.17\times 43.4$~cm$^{3}$) of plastic scintillator which are placed vertically between the $E$-scintillators and the target. All $E$- and $\Delta E$-scintillators are made of BICRON-408 plastic scintillator material. Due to energy losses in materials between the target and the $E$-scintillators, the protons (deuterons) with an initial energy below $20$~MeV ($25$~MeV) will not reach the $E$-scintillators.

The target system of BINA~\cite{Nasser98} consists of a target cell, a holder, a cryogenic system, a heater, a gas-flow system, temperature sensors, and a temperature controller unit. We used deuterium (LD$_2$) with density of $\rho=169$~mg/cm$^{3}$. The effective target thickness was 3.85$\pm$0.2 mm, including the bulging effect. The target cell used in this experiment was made of high purity Aluminum to optimize the thermal conductivity and its windows were covered by a transparent foil of Aramid with a thickness of 4~$\mu$m. The operating temperature and pressure of the LD$_2$ target were 19~K and 258~mbar, respectively. The target holder was installed at $\theta_{lab}=100^{\circ}$ on top of the backward ball with a slight inclination angle of $10^{\circ}$ and could be moved by a pneumatic system.

The backward part of BINA is ball-shaped and is made out of 149 phoswich detectors. These detectors cover almost 80\% of the full $4\pi$ solid angle, polar angle, $\theta$, in the range of $40^{\circ}$ to $165^{\circ}$ with a complete azimuthal acceptance ($\phi$) (except at the position of the target holder at $\theta=100^{\circ}$). Therefore, the backward ball together with the forward wall cover nearly the complete phase space. The shape and the construction of the inner surface of the ball is similar to the surface of a soccer ball (which consists of 20 identical hexagon and 12 identical pentagon structures). Each pentagon (hexagon) is composed of five (six) identical triangles. In the hexagon, all sides of the triangle have the same size while in the pentagon only two sides are the same. Each triangle is composed of a phoswich detector and covers an angular range as large as $\sim20^{\circ}$, in both $\phi$ and $\theta$ directions. Therefore, the granularity of the backward ball is poor compared to that of the forward wall. Each detector of the backward ball is composed of a fast plastic scintillator, BICRON BC-408, and a slow phoswich part, BICRON BC-444, which has the same cross section and is glued to the fast component. The slow scintillator part has a thickness of 1~mm, while, because of the energy difference between particles scattered at different polar angles, the thickness of the fast scintillator below $\theta<100^{\circ}$ is 9~cm and for the rest is 3~cm. All these elements were glued with each other making a spherical ball. More details of the backward ball can be found in Ref.~\cite{Mohammad_thesis}.

The front exit window of the backward ball was made of 250~$\mu$m thick Kevlar cloth and 50~$\mu$m thick Aramica foil~\cite{Nasser98} which are glued to a metal frame. This thin window is strong enough to hold the vacuum inside the ball (with a pressure of $10^{-5}$ mbar) and it also allows the forward scattered particles to pass through it with a very small energy loss.

The BINA backward ball acts as a scattering chamber. The achieved vacuum is sufficient to avoid the collection of dirt on the foil of the liquid-deuterium target. With such an active scattering chamber, scattered particles lose less energy compared to those propagating to the forward wall. Therefore, the ball detects particles with low kinetic energies. The energy threshold is, in the ball case, determined by the material related to the target cell, such as the target frame, the target window foil and the thin cylindrical aluminum foil used as a thermal shielding around the target cell.

The electronic, read-out and data acquisition (DAQ) systems were adapted from the former SALAD setup~\cite{Ela10}. Four different trigger conditions were used in this work. These conditions were based on hit multiplicity in photo-multiplier tubes (PMTs) of $E$- and $\Delta E$-scintillators and ball detectors. A Faraday cup at the end of the beam was used for stopping the beam and monitoring its current. A precision current meter was connected to the Faraday cup. The output of the current meter was converted into logic signals with a frequency proportional to the actual current and read out by the scalers of the DAQ. The beam current was typically 15~pA.
\section{Data analysis}
\label{Analysis}
In this section, the analysis of the proton-deuteron break-up reaction for the forward-backward configurations will be discussed. A thorough description of the data analysis of the forward-forward configurations can be found in Refs.~\cite{Hajar_2020,Hajar_2019,Hajar_FB22_2020}.
\subsection{Events selection and energy calibration}
\label{selection}
Events were selected for which two break-up proton candidates were found in coincidence in the final state. In this work, the forward-backward configurations in which one of the outgoing protons scatters to the forward wall and the second one to the backward part of the setup were selected. The angular bins for event integration were chosen to be $\Delta\theta_{1}=20^{\circ}$ (size of the ball detector), $\Delta\theta_{2}=4^{\circ}$ and $\Delta\phi=10^{\circ}$.

Energy calibration is done using the break-up channel itself and exploiting the energy correlation between the two protons in the final state. For translating the Charge-to-Digital Converter (QDC) channel into the deposited energy by a particle, we need to know the energy correlation of the two protons at the detector position based on their scattering angles and energy losses. We decided to convert the theoretical kinematic curve at the target position into the one at the detector position. This conversion is done by determining the energy loss due to the materials between the target and the detector using GEANT3~\cite{Geant4_2006} simulations.

\begin{figure}[!ht]
\centering
\includegraphics[angle=0,width=0.49\textwidth]{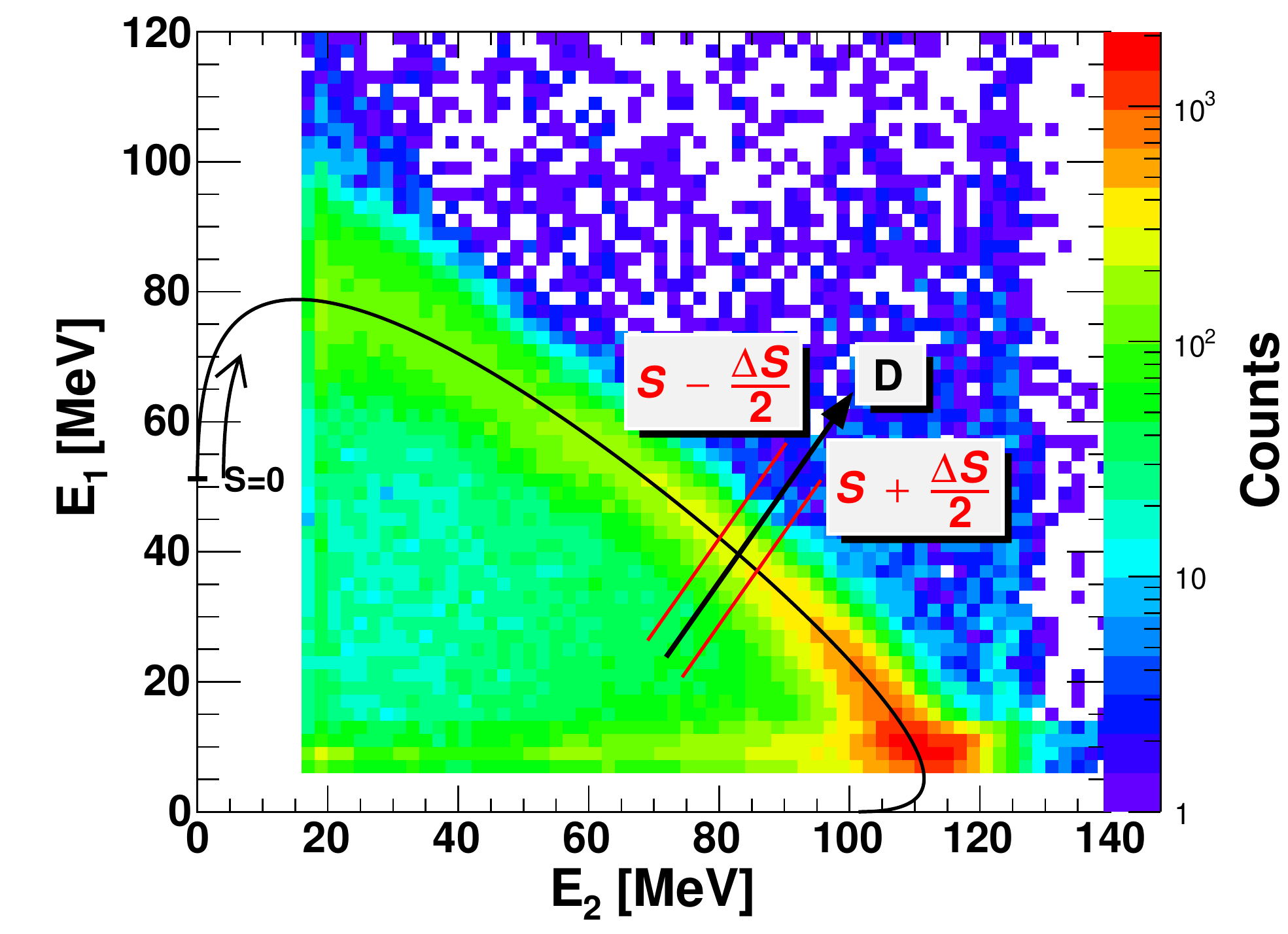}
\caption{The energy spectrum of two coincident protons coming from the break-up reaction
 and registered at ($\theta_{1}=50^\circ \pm 10^\circ$, $\theta_{2}=28^\circ \pm 2^\circ$,
 $\phi_{12}=140^\circ \pm 5^\circ$). The solid line shows the kinematical $S$-curve for the central
 values of the experimental angular bins. The starting point of $S=0$ is 
indicated by a small bar with the appropriate label. The value of $S$ increases in the 
direction of the arrow presented near $S=0$. The red lines indicate a selected window 
corresponding to a mean value of $S$ of 135~MeV with $\Delta S=10$~MeV.}
\label{p1p2thr_50_28_140_Daxis}
\end{figure}

The break-up observables are shown as a function of $S$, the arc length along the $S$-curve. The $S$-curve is the kinematical curve presenting the energy correlation between two final-state particles of the break-up reaction. The energy losses were added to the deposited energies to convert them into initial energies at the interaction point or target position and one of the results is presented in Fig.~\ref{p1p2thr_50_28_140_Daxis} for the configuration with $\theta_{1}=50^\circ$, $\theta_{2}=28^\circ$ and $\phi_{12}=140^\circ$. This configuration corresponds to break-up protons scattered to a subring which consists of 5 ball elements with their centroids placed at a common polar angle of $50^\circ$ in coincidence with protons detected at $28^\circ\pm2^{\circ}$ in the forward wall. The finite width of the band is predominantly determined by the large angular coverage of each ball element. Therefore, several sub-configurations fall within the acceptance of the detector; see Ref.~\cite{Maisam_2020}. The energy threshold for registering protons in the wall (ball) is about $15$~MeV ($7$~MeV).
\begin{figure}[!ht]
\centering
\includegraphics[angle=0,width=0.49\textwidth]{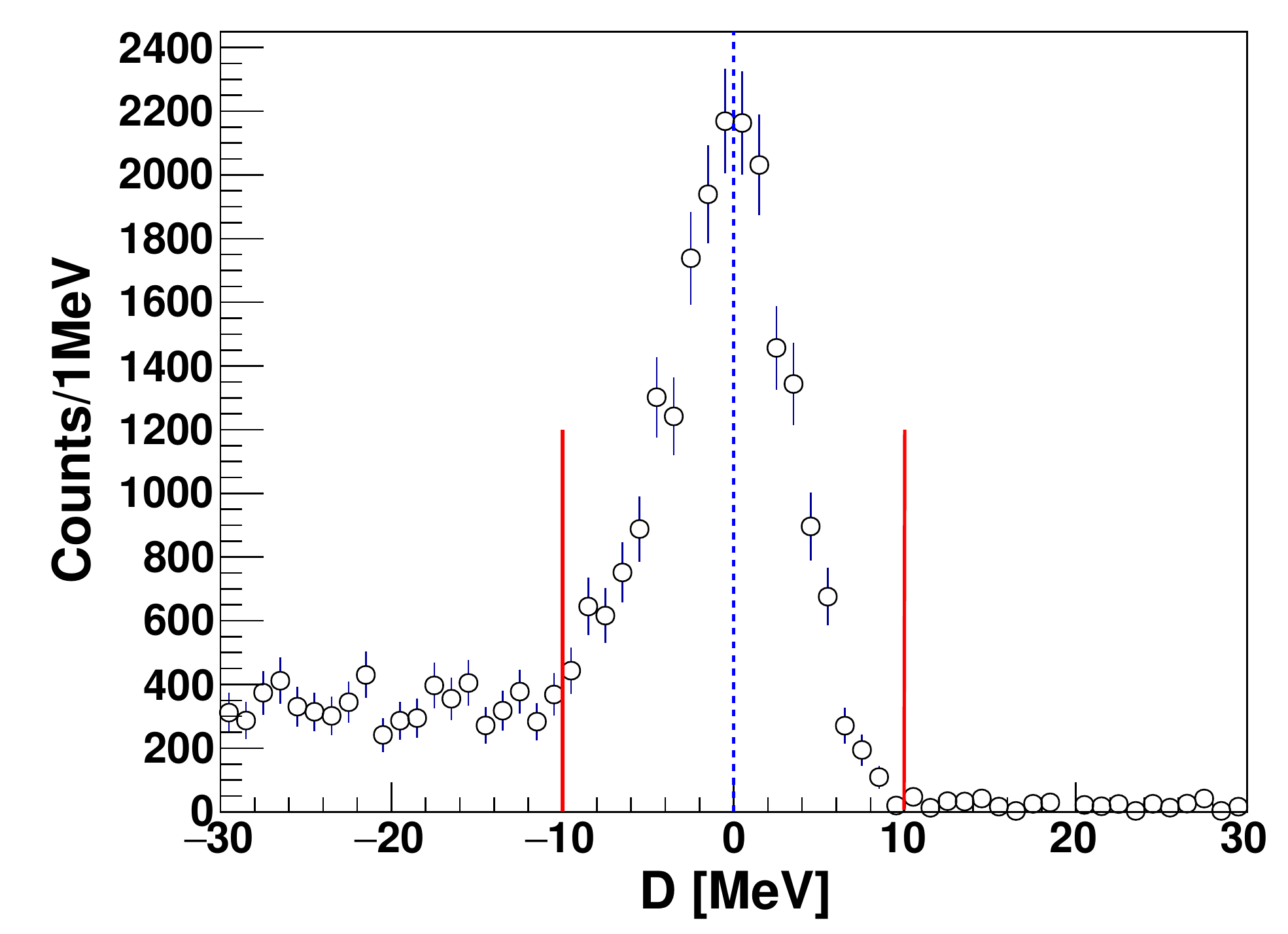}
\caption{The projection of the slice chosen in Fig.~\ref{p1p2thr_50_28_140_Daxis} on the D-axis for $S=135$~MeV.
The vertical red lines mark the selection window corresponding to $\pm 3\sigma$ around the peak position.}
\label{Daxis}
\end{figure}
\subsection{Determination of the analyzing powers}
\label{AnalyzingPowers}
To obtain the analyzing powers as a function of $S$, the $S$-curve is divided into slices ($S$-bin) of equal width of $\Delta S$ ($10$~MeV) along its length; see Fig.~\ref{p1p2thr_50_28_140_Daxis}. The projection of the indicated region in Fig.~\ref{p1p2thr_50_28_140_Daxis} onto the $D$-axis (a line perpendicular to the $S$-curve) is shown in Fig.~\ref{Daxis}. Candidate signal events are selected within $\pm$10~MeV of $D$ corresponding to $\pm$3$\sigma$ around the observed peak position. The number of events is normalized to the collected charge corrected for the dead time.

The general formula for the cross section of the break-up reaction induced by an incident polarized beam made up of spin-$\frac{1}{2}$ particles in the Cartesian coordinate system is given by Ref.~\cite{Ohlsen81}:
\begin{eqnarray} 
\sigma(\xi,\phi_{12})=\sigma_{0}(\xi,\phi_{12})[1&+&p_{x}A_{x}(\xi,\phi_{12})\nonumber\\
&+&p_{y} A_{y}(\xi,\phi_{12})\nonumber\\
&+&p_{z} A_{z}(\xi,\phi_{12})],
\label{cross1}
\end{eqnarray}
where $\sigma_{0}$ is the cross section for the case of an unpolarized beam, $p_{x}$, $p_{y}$ and $p_{z}$ are the Cartesian components of the beam polarization, $A_{x}$, $A_{y}$ and $A_{z}$ refer to the analyzing powers, and $\phi_{12}=\phi_{1}-\phi_{2}$ together with $\xi$=($\theta_{1},\theta_{2},S$) denote all the kinematical variables of the two outgoing particles in the break-up reaction. Components of the beam polarization are related to the beam polarization $p_{Z}$ with respect to the quantization axis~\cite{Ohlsen81}. Making use of these relations for the beam polarization normal to its momentum and following Eq.~\ref{cross1}, the $\phi$ dependence of the number of events $N_{\xi,\phi_{12}}^{\uparrow}$ ($N_{\xi,\phi_{12}}^{\downarrow}$) for the spin-up state, $\uparrow$ (spin-down state, $\downarrow$) and for a kinematical point ($\xi,\phi_{12}$) can be written as:
\begin{eqnarray} 
N_{\xi,\phi_{12}}^{\uparrow,\downarrow}(\phi)=N_{\xi,\phi_{12}}^{0}(\phi)[1&-&p_{Z}^{\uparrow,\downarrow}A_{x}(\xi,\phi_{12})\sin\phi\nonumber\\
&+&p_{Z}^{\uparrow,\downarrow}A_{y}(\xi,\phi_{12})\cos\phi],
\label{Anaequation1}
\end{eqnarray}
where $\xi$=($\theta_{1},\theta_{2},S$) denotes all the kinematical variables except $\phi_{12}$, and $p_{Z}^{\uparrow}$ and $p_{Z}^{\downarrow}$ are the polarization of the up and down polarized beam, respectively, with respect to its quantization axis. $N_{\xi,\phi_{12}}^{0}$ is the number of events for the case of an unpolarized beam. According to Eq.~\ref{Anaequation1}, by eliminating the $N_{\xi,\phi_{12}}^{0}(\phi)$ term, the following formula is obtained.
\begin{eqnarray} 
\frac{N_{\xi,\phi_{12}}^{\uparrow}(\phi)-N_{\xi,\phi_{12}}^{\downarrow}(\phi)}{p_{Z}^{\uparrow}N_{\xi,\phi_{12}}^{\downarrow}(\phi)-p_{Z}^{\downarrow}N_{\xi,\phi_{12}}^{\uparrow}(\phi)}=&-&A_{x}(\xi,\phi_{12})\sin\phi\nonumber\\
&+&A_{y}(\xi,\phi_{12})\cos\phi.
\label{Anaequation2}
\end{eqnarray}
\begin{figure}[!ht]
\centering
\includegraphics[angle=0,width=0.49\textwidth]{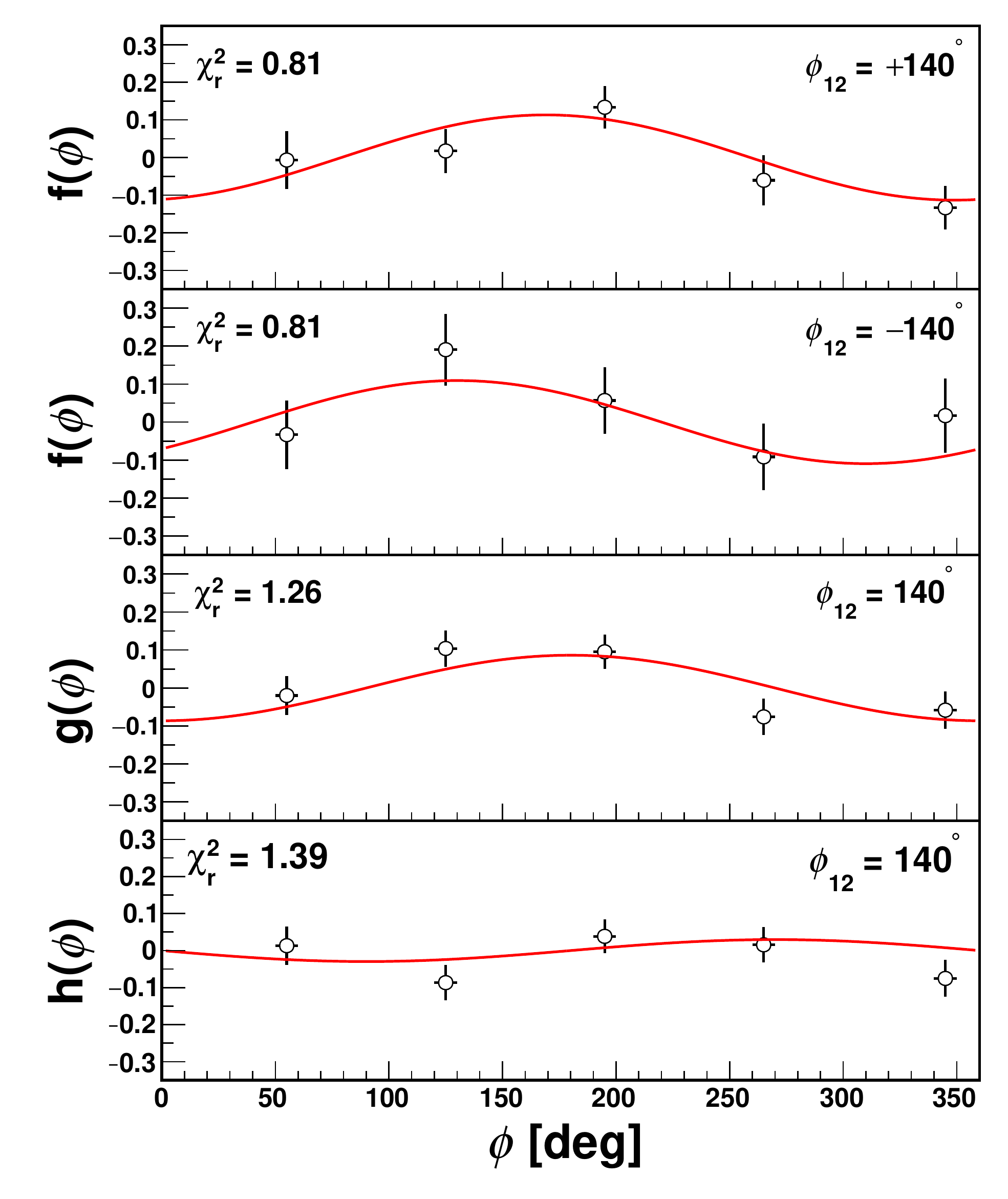}
\caption{Examples of asymmetry distributions with fitted curves obtained for configurations 
with $\theta_{1}=50^\circ$, $\theta_{2}=28^\circ$, $\phi_{12}=\pm140^\circ$ and $S=135$~MeV. 
The two upper (lower) panels depict the asymmetries $f_{\xi,+\phi_{12}}(\phi)$ 
and $f_{\xi,-\phi_{12}}(\phi)$ ($g_{\xi,\phi_{12}}(\phi)$ and $h_{\xi,\phi_{12}}(\phi)$). 
The data are represented as open circles and the red lines show the results of a fit through 
the data. See text for further details.}
\label{Asymmetry_k2bin4_50_28_140}
\end{figure}
\noindent By denoting the left side of Eq.~\ref{Anaequation2} by $f_{\xi,\phi_{12}}(\phi)$, one can rewrite the Eq.~\ref{Anaequation2} as follows:
\begin{equation} 
f_{\xi,\phi_{12}}(\phi)=-A_{x}(\xi,\phi_{12})\sin\phi+A_{y}(\xi,\phi_{12})\cos\phi.
\label{Anaequation3}
\end{equation}
\begin{figure}[!t]
\centering
\includegraphics[angle=0,width=0.50\textwidth]{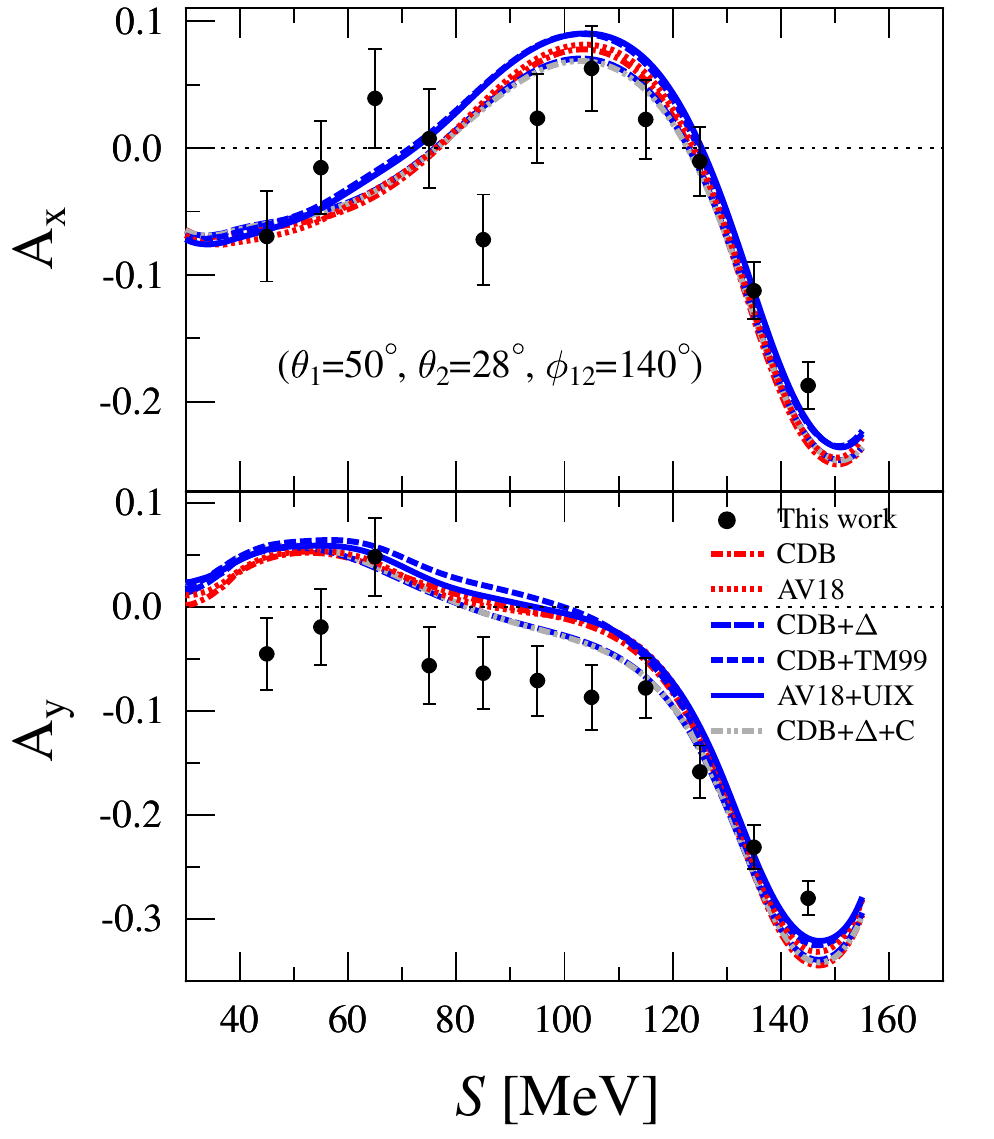} 
\caption{Examples of the analyzing powers $A_{x}$ and $A_{y}$ of the 
proton-deuteron break-up reaction for one kinematical configuration 
($\theta_{1}=50^\circ$, $\theta_{2}=28^\circ$, $\phi _{12}=140^\circ$). 
Theoretical predictions, as specified in the legend, show the Faddeev 
calculations using the 2NF such as CDB~\cite{Machleidt96,Machleidt01} (dashed-dotted line) and 
AV18~\cite{Wiringa95} (dotted line) and 2NF+3NF models such as CDB+$\Delta$ (long-dashed line), 
CDB+TM99~\cite{Witala98,Witala01,Witala2009} (short-dashed line), AV18+UIX~\cite{Deltuva_09} (solid line) 
and CDB+$\Delta$+Coulomb~\cite{Deltuva2005b,Deltuva2006} (dashed-double-dotted line).}
\label{Ax_Ay_50_28_140}
\end{figure}
\noindent Thus, $A_{x}$ and $A_{y}$ values can be extracted, if one uses the following combinations for mirror configurations ($\xi,+\phi_{12}$) and ($\xi,-\phi_{12}$):
\begin{equation} 
g_{\xi,\phi_{12}}(\phi)\equiv \frac{f_{\xi,-\phi_{12}}(\phi)+f_{\xi,+\phi_{12}}(\phi)}{2},
\label{Anaequation4}
\end{equation}
and 
\begin{equation} 
h_{\xi,\phi_{12}}(\phi)\equiv \frac{f_{\xi,-\phi_{12}}(\phi)-f_{\xi,+\phi_{12}}(\phi)}{2},
\label{Anaequation5}
\end{equation}
which, using parity conservation, can be expressed as:
\begin{equation} 
g_{\xi,\phi_{12}}(\phi)=A_{y}\cos\phi,
\label{Anaequation6}
\end{equation}
and 
\begin{equation} 
h_{\xi,\phi_{12}}(\phi)=A_{x}\sin\phi.
\label{Anaequation7}
\end{equation}
\noindent Using the beam polarizations $p_{Z}^{\uparrow}=0.57\pm0.02$ and $p_{Z}^{\downarrow}=-0.70\pm0.04$ from Ref.~\cite{ahm1}, $A_{y}$ ($A_{x}$) was extracted by fitting the experimentally determined distribution $g_{\xi,\phi_{12}}(\phi)$ ($h_{\xi,\phi_{12}}(\phi)$) with the right-hand side function of Eq. \ref{Anaequation6} (Eq. \ref{Anaequation7}). Samples of such fits for a particular $S$-bin in a given configuration are illustrated in Fig.~\ref{Asymmetry_k2bin4_50_28_140}. The extraction of the analyzing powers relies on determining ratios of normalized rates measured with up and down polarized beams. Therefore, many experimental factors like detection efficiency of MWPC and scintillators, and uncertainties in the determination of the solid angles cancel. Figure~\ref{Ax_Ay_50_28_140} shows the two extracted analyzing powers, $A_{x}$ and $A_{y}$, for one of chosen kinematical configurations. The error bars reflect only statistical uncertainties.
\subsection{Error analysis}
\label{ErrorAnalysis}
This section describes the procedure that has been used to extract statistical and systematical uncertainties. We give an overview of the various sources of systematic uncertainties that have been identified and discuss the methodology that has been used to estimate their magnitudes.

The statistical uncertainties for the analyzing powers $A_{y}$ and $A_{x}$ arise from the errors of fitting parameters of the functions $A_{y}\cos\phi$ and $A_{x}\sin\phi$ fitted to $g_{\xi,\phi_{12}}(\phi)$ and $h_{\xi,\phi_{12}}(\phi)$, respectively. Statistical uncertainties of $g_{\xi,\phi_{12}}(\phi)$ and $h_{\xi,\phi_{12}}(\phi)$ were obtained by performing the error propagation in Eq.~\ref{Anaequation4} and Eq.~\ref{Anaequation5}, i.e.,
\begin{eqnarray} 
\Delta g_{\xi,\phi_{12}}(\phi)&=&\Delta h_{\xi,\phi_{12}}(\phi)\nonumber\\
&=&\frac{1}{2}\sqrt{\Big(\Delta f_{\xi,+\phi_{12}}(\phi)\Big)^{2}+\Big(\Delta f_{\xi,-\phi_{12}}(\phi)\Big)^{2}}, 
\label{AP_Error}
\end{eqnarray}
\noindent where the statistical uncertainty in the functions $f_{\xi,+\phi_{12}}(\phi)$ and $f_{\xi,-\phi_{12}}(\phi)$ was obtained by performing the error propagation in the left side of Eq.~\ref{Anaequation2}, i.e.,
\begin{eqnarray}
\left.\begin{aligned}
\Delta f_{\xi,\pm\phi_{12}}(\phi)=\nonumber\\
\end{aligned}\right.
\frac{(p_{Z}^{\uparrow}-p_{Z}^{\downarrow})}{\Big(p_{Z}^{\uparrow}N_{\xi,\pm\phi_{12}}^{\downarrow}(\phi)-p_{Z}^{\downarrow}N_{\xi,\pm\phi_{12}}^{\uparrow}(\phi)\Big)^{2}}\nonumber\\
\times\sqrt{(N_{\xi,\pm\phi_{12}}^{\uparrow})^{2}N_{\xi,\pm\phi_{12}}^{\downarrow}+(N_{\xi,\pm\phi_{12}}^{\downarrow})^{2}N_{\xi,\pm\phi_{12}}^{\uparrow}}.
\label{f_rror} 
\end{eqnarray}

For the analyzing powers, one of the contributions to the systematic uncertainty, which does not cancel in the ratios given by Eq.~\ref{Anaequation2}, stems from the uncertainty of the beam polarizations. The estimated values of uncertainty related to this effect were $\sim$3\% and $\sim$6\% for the up and down-modes, respectively~\cite{ahm1}. Altogether, by adding these two systematic uncertainties in quadrature, the maximum systematic uncertainty associated with this effect for analyzing powers is estimated to be less than 7\%.

In addition to the systematic error due to the uncertainty in the beam polarization, we considered other sources of uncertainty that stem from residual and unknown asymmetries. Some of the asymmetries might be caused by variations in the efficiency and beam currents between the data taken with the up and down polarization states. Moreover, very small differences between the position of the beam-target interaction point between the two polarization states have been considered as a source of systematic uncertainty. During data taking we minimized these effects by regularly monitoring the position of the interaction point via light intensity measurements of the beam impinging a ZnS target. No deviations were visually observed implying variations that are less than 1~mm. All possible residual asymmetries not related to the analyzing powers have been estimated by applying a different fit function to the one presented in Eqs.~\ref{Anaequation6} and~\ref{Anaequation7}. For this purpose, the analyzing powers are measured by fitting $g_{\xi,\phi_{12}}(\phi)$ and $h_{\xi,\phi_{12}}(\phi)$ to the functions $A_{y}\cos\phi+A$ and $A_{x}\sin\phi+B$, respectively. The magnitude of this systematic uncertainty was estimated by taking the difference between the analyzing powers with and without the free parameters ($A$ and $B$) of the fitting functions. The typical uncertainty related to this effect, on the final analyzing powers $A_x$ and $A_y$, was found to be around 0.015 and 0.005, respectively.

The analyzing powers $A_x$ and $A_y$ were extracted by selecting events that fall within $3\sigma$ around the peak position in the $D$-spectrum; see Fig.~\ref{Daxis}. We note that most of the events on the left-hand side of the peak stem from break-up events whereby one of the protons undergo a hadronic interaction in the scintillator material. Therefore, only a small fraction of events that fall within the selection window is due to background. To estimate the effect of the residual background on $A_x$ and $A_y$, we performed an alternative analysis procedure. For this, we extracted the analyzing powers for data that fall within the interval $-3\sigma$ and $0$ of $D$ and for the interval starting from 0 to $+3\sigma$. The difference between these two data samples we used as an estimate of the systematic uncertainty due to the background. The resulting analyzing powers for the left and right sides differ at most by 0.01 for both $A_{y}$ and $A_{x}$. The total systematic uncertainty is obtained by adding all contributions in quadrature assuming them to be independent.
\section{Theoretical calculations}
\label{Theory}
\textcolor{referee1}{Theoretical predictions of the present work are obtained within rigorous frameworks that are based on only pairwise 2N interactions or based on a combination of both 2NF and 3NF in the nuclear Hamiltonian. The 2NF, the so-called realistic potentials, contain commonly a local one-pion exchange potential (OPEP) part to account for the long-range NN interaction, but differ in their short and intermediate-range parts which are generally non-local. We employ the following realistic NN potentials: CDB~\cite{Machleidt96,Machleidt01} and AV18~\cite{Wiringa95}. These potentials can be combined with 3NF models, which are refined versions of the 3NF proposed originally by Fujita and Miyazawa~\cite{Fujita57} to describe a system composed of more than two nucleons.}

\textcolor{referee1}{Specifically, we apply first the formalism of the Faddeev equations to obtain predictions based on the two-nucleon CDB or AV18 interactions only. Next, we extend our treatment of nuclear interaction and combine these 2NFs with the TM99~\cite{Coon79,Coon01} or the UIX~\cite{Carlson83,Pudliner95} 3N potentials, respectively. We also apply the coupled-channel approach in which in addition to the CDB interaction we take into account the explicit $\Delta$-isobar excitations. Within this approach the Coulomb interaction between protons is also included.}

\textcolor{referee1}{In our simplest theoretical approach only two-body interactions $V_{ij}$ contribute to the 3N Hamiltonian. In such a case the transition amplitude for the deuteron break-up 
\begin{equation}
U_0 = (1+P)T
\label{eq_U0}
\end{equation}
is given in terms of the break-up operator $T$ satisfying the Faddeev-type integral equation~\cite{Glockle96}
\begin{equation}
T \vert \psi \rangle = t P \vert \psi \rangle + t P G_0 T \vert \psi \rangle.
\label{eq_Faddeev_2N}
\end{equation}
The initial state $\vert \psi \rangle$ is a product of the internal deuteron state and the relative nucleon-deuteron momentum state. Further, the off-shell two-nucleon t-matrix $t$ results from the pairwise interaction $V_{23}$ (in one selected 2N sub-system) through the 2N Lippmann-Schwinger equation, and $G_0$ is the free 3N propagator. Finally, the permutation operator $P = P_{12}P_{23} + P_{13}P_{23}$ is given in terms of transpositions $P_{ij}$ which interchange nucleons $i$ and $j$. The physical picture underlying Eq.~\ref{eq_Faddeev_2N} is revealed by its iterations which yield a multiple scattering series for $T$.}

\textcolor{referee1}{The second group of presented predictions arises from including, in addition to the 2N interaction $V_{ij}$, also three-nucleon force $V_4 \equiv V_{123}$. Taking advantage of the fact that each 3N interaction can be split in three parts $V_4^{(i)}$ which are symmetrical under exchanges of nucleons $j \neq i$ and $k \neq i$,
the Faddeev equation for the break-up operator $T$ is expressed as~\cite{Huber97}
\begin{eqnarray}
T \vert \psi \rangle &=& tP \vert \psi \rangle + (1+tG_0)V_4^{(1)}(1+P)\vert \psi \rangle \nonumber \\
&+& t P G_0 T \vert \psi \rangle + (1+tG_0)V_4^{(1)}(1+P)G_0 T\vert \psi \rangle,
\label{eq_Faddeev_3N}
\end{eqnarray}
while the transition amplitude $U_0$ remains as in Eq.~\ref{eq_U0}.}
 
\textcolor{referee1}{The numerical methods used to solve Eqs.~\ref{eq_Faddeev_2N} and~\ref{eq_Faddeev_3N} are discussed in detail in Refs.~\cite{Glockle96,Huber97,Glockle_book}. In short, we work in momentum space and build the 3N partial wave basis from the Jacobi relative momenta and a set of discrete quantum numbers describing orbital angular momenta, spins and isospins in the 3N system. Next we project Eqs.~\ref{eq_Faddeev_2N} and~\ref{eq_Faddeev_3N} onto state basis, what leads to a finite set of coupled integral equations with two continuous variables. We solve it iteratively, generating a Neumann series which we sum up by the Pad\'e method. Once the matrix elements of $T$ are known the transition amplitudes $U_0$ and observables are computed. In the case of the TM99 3NF its free cut-off parameter $\Lambda$ was adjusted so that this force in combination with the CDB NN potential reproduced the experimental triton binding energy~\cite{Witala01}.}

\textcolor{referee1}{Alternative approach to study the 3N break-up cross section relies on the symmetrized Alt-Grassberger-Sandhas (AGS) form of Faddeev equations~\cite{Deltuva2003a}. As shown in Refs.~\cite{Deltuva2005b,Deltuva2006} the three-particle break-up matrix $U_0^{(R)}$, 
formally depending also on the screening radius $R$, fulfills
\begin{equation}
U_0^{(R)} = (1 + P) G_0^{-1} + (1 + P)T^{(R)} G_0 U^{(R)},
\end{equation}
where $T^{(R)}$ is the two-particle transition matrix derived from nuclear plus screened Coulomb potentials and 
\begin{equation}
U^{(R)} = P G_0^{-1} + P T^{(R)} G_0 U^{(R)}
\end{equation}
is the AGS three-body transition operator. Working in this formalism we use the two-nucleon coupled-channel potential~\cite{Deltuva2003b} which includes states in which one nucleon is turned into a $\Delta$ isobar. The presence of the $\Delta$ isobar generates an effective 3N force.} 

\textcolor{referee1}{In the practical computations the Neumann series for the on-shell matrix elements of the operator $U_0^{(R)}$ is obtained and summed up by the Pad\'e method. The approach given in Ref.~\cite{Deltuva2006} allows us to include efficiently the Coulomb interaction omitted in the Faddeev equation-based formalism described above.}

\textcolor{referee1}{Summarizing, in following sections we show predictions based on the 2NF potentials (the CDB or the AV18) or on the 2N+3N forces: CDB+TM99~\cite{Witala98,Witala01,Witala2009}, AV18+UIX~\cite{Witala2009} obtained within the Faddeev approach and CDB+$\Delta$~\cite{Deltuva2003a,Deltuva2003b} and CDB+$\Delta$+Coulomb~\cite{Deltuva2005b,Deltuva2006} results from the AGS scheme. To compare these theoretical predictions with the experimental data, they are all averaged over the detector acceptances. Below, the averaging procedure is briefly outlined.}
\subsection{Averaging of the theoretical predictions over experimental acceptance}
\label{Averaging}
As explained in Sec.~\ref{Exp}, each ball detector covers a large solid angle. Therefore, the experimentally-extracted observables are integrated over a large part of the solid angle, and, hence, one cannot simply assume that the results correspond to those measured at the central coordinate of the detector. Thus, in order to perform a fair comparison between the data and the results of the calculations, averaging~\cite{Ela10} of the theoretical values of the observables over the experimental detector acceptance has been applied.
\begin{figure}[!ht] 
\centering
\includegraphics[angle=0,width=0.49\textwidth]{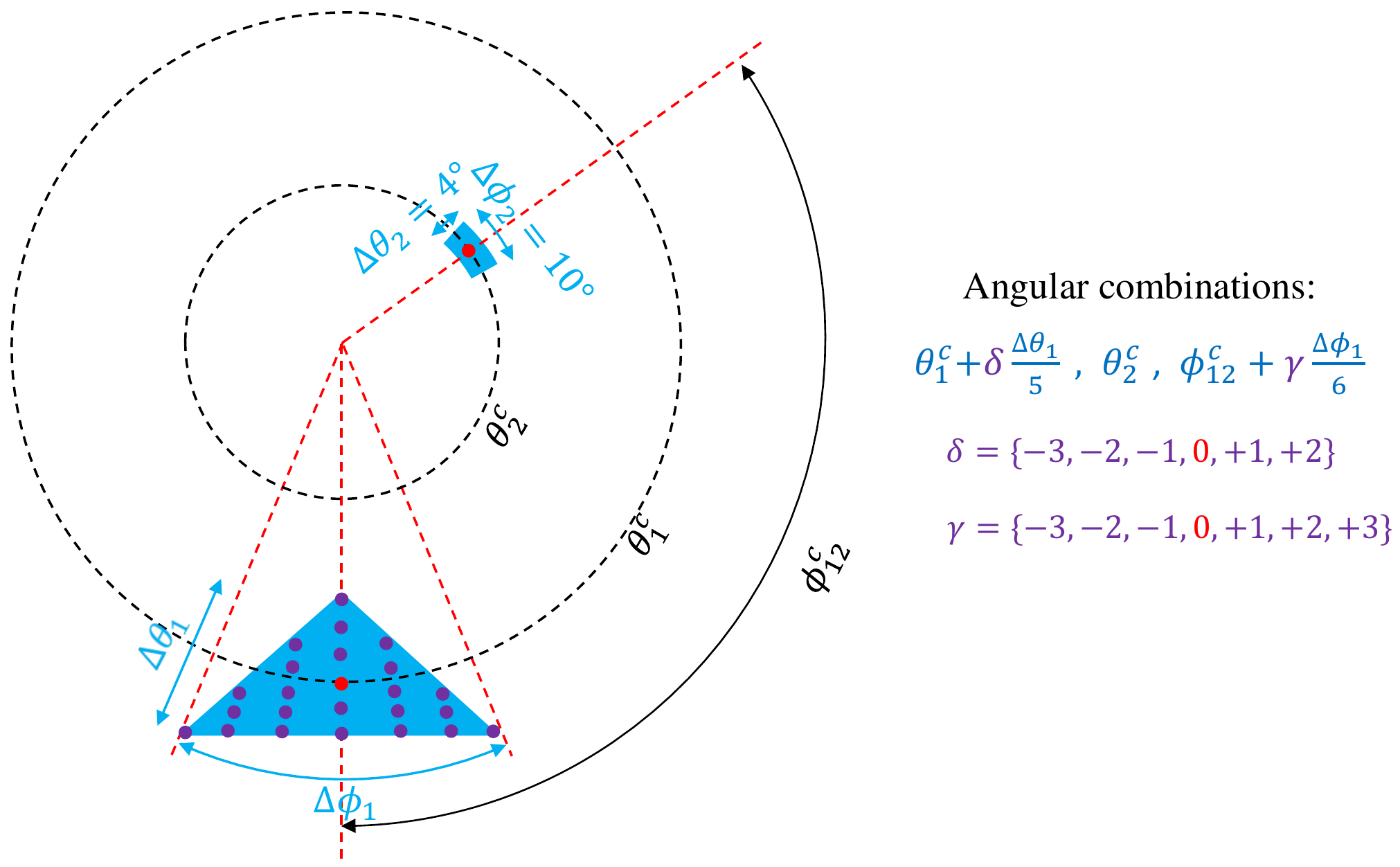}
\caption{Schematic drawing of the angular bins used for event integration at 
$(\theta_{1}=50^{\circ},\theta_{2}=28^{\circ})$. Indices ``1'' and ``2'' represent 
the backward and forward angles, respectively. The central configuration 
($\theta_{1}^{c}$, $\theta_{2}^{c}$, $\phi_{12}^{c}$) is marked with red and other 
combinations of angles are shown by purple with $\delta$ and $\gamma$ taking values 
specified in the legend. Note that the numbers in the 
legend are specific for the detector shown here.} 
\label{Angular_bins_r1s2}
\end{figure}

Figure~\ref{Angular_bins_r1s2} shows the schematic drawing of the backward and forward angular bins that are used to count exclusively the break-up events. First, for each configuration defined by the central values of angles $\theta_{1}^{c}$, $\theta_{2}^{c}$ and $\phi_{12}^{c}$, the theoretically-determined analyzing powers ($A_{x}$ and $A_{y}$) and cross sections ($\sigma$) are obtained for all combinations of angles $\theta_{1}^{c}+\delta\frac{\Delta\theta_{1}}{5}$, $\theta_{2}^{c}$ and $\phi_{12}^{c}+\gamma\frac{\Delta\phi_{1}}{6}$, where $\delta$ and $\gamma$ are integer numbers (specified in the legend). Only those combinations that fall within the acceptance of the detector are considered without taking into account the size of the forward angular bins; see Fig.~\ref{Angular_bins_r1s2}. Then, analyzing-power values are weighted with the product of the 5-fold differential cross section for that value of the angle and the solid angle factor, while the cross section values are only weighted with the solid angle factor. Finally, the weighted observables are placed on the $E_{1}$ versus $E_{2}$ plane to project them onto the relativistic $S$-curve calculated for the central angles $\theta_{1}^{c}$, $\theta_{2}^{c}$ and $\phi_{12}^{c}$. In this way, the results of the non-relativistic calculations are projected onto relativistic kinematics and, therefore, non-relativistic calculations can be directly compared to the $S$ distributions of the data, without the necessity to correct for difference of arc-lengths calculated along relativistic and non-relativistic $S$-curves. Note that in this step, the variable $S$ was not used as a reference point for the configurations because $S$ is defined individually for each of them, therefore, the same values of $S$ for different configurations correspond usually to different ($E_{1}$,$E_{2}$) points. This is merely the consequence of a large-size detector containing many kinematical configurations. 
\section{Experimental results}
\label{results}
Experimental results of the analyzing powers ($A_x$ and $A_y$) for 105 kinematical configurations are given in the supplementary material. In Fig.~\ref{Ax_Ay_45_24_for_Paper}, the analyzing powers at $(\theta_{1}=45^{\circ},\theta_{2}=24^{\circ})$ as a function of $S$ are presented for different azimuthal opening angles. Error bars reflect only statistical uncertainties and the cyan bands show the systematic uncertainties. In this figure, one can see that in general for a given configuration in the whole range of $S$, the data lie systematically above, on, or below the theoretical predictions. For instance, for the analyzing powers $A_y$ at $\phi_{12}=20^{\circ}$, data lie above and, towards $\phi_{12}=180^{\circ}$, the data are located below the theory predictions. The agreement between data and theoretical calculations depends strongly on $\phi_{12}$ and less on the variable $S$. Therefore, we decided to integrate the observables over $S$ that facilitates the comparison with the calculations. In this method, both measured and calculated data points of the analyzing powers ($A_{x}$ and $A_{y}$) for each configuration, $(\theta_{1},\theta_{2},\phi_{12})$, are averaged over $S$ using the following equation:
\begin{eqnarray}
\bar{A}_{x(y)}(\xi)=
\dfrac{\sum \limits_{i=1}^{N} \dfrac{A_{x(y)}(\xi,S_{i})}{(\Delta A_{x(y)}^{exp}(\xi,S_{i}))^{2}}}
{\sum \limits_{i=1}^{N} \dfrac{1}{(\Delta A_{x(y)}^{exp}(\xi,S_{i}))^{2}}},
\label{ave_equation}
\end{eqnarray}
where $\xi$=($\theta_{1},\theta_{2},\phi_{12}$) denotes all kinematical variables excluding $S_{i}$. $N$ is the number of data points in $S$ for that configuration, $\Delta A_{x(y)}(\xi,S_{i})$ is the uncertainty of the data point and $i$ is the index for the variable $S$ running from 1 to $N$. The uncertainty in the experimental average can be evaluated using standard error propagation as
\begin{eqnarray}
\Delta \bar{A}_{x(y)}(\xi)=
\dfrac{1}{\sqrt{\sum \limits_{i=1}^{N} \dfrac{1}{(\Delta A_{x(y)}^{exp}(\xi,S_{i}))^{2}}}}.
\label{err_ave_equation}
\end{eqnarray}
Figures~\ref{Ax_weighted_average_bw_wwHajar} and~\ref{Ay_weighted_average_bw_wwHajar} present the averages of the analyzing powers $A_{x}$ and $A_{y}$, respectively, as a function of the opening azimuthal angle, $\phi_{12}$. The errors are statistical and the cyan bands depict $2\sigma$ systematic uncertainties. The averages of the calculations are presented by the same line colors and styles as were chosen for Fig.~\ref{Ax_Ay_50_28_140}.
\hvFloat[
 floatPos=!htp,
 capWidth=w,
 capPos=b,
 capAngle=0,
 objectAngle=0,
 capVPos=t,
 objectPos=t]
{figure*}{\includegraphics[angle=0,width=0.85\textwidth]{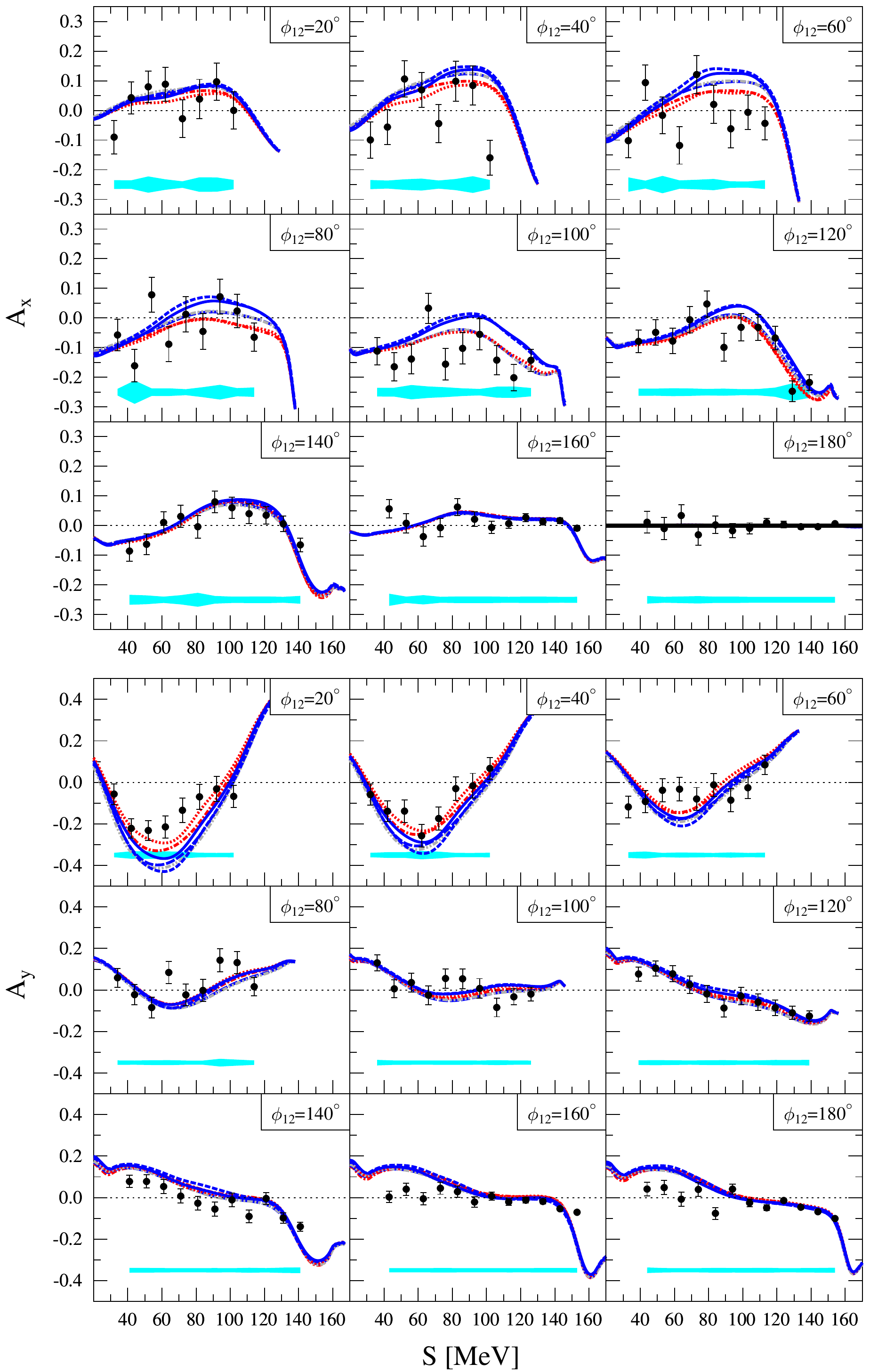}}
{The analyzing powers at $(\theta_{1}=45^{\circ},\theta_{2}=24^{\circ})$ 
as a function of $S$ for different azimuthal opening angles. Error bars 
reflect only statistical uncertainties. The cyan bands show the total systematic 
uncertainty whereby the width corresponds to 2$\sigma$. For a description of the 
lines, we refer to the caption of Fig.~\ref{Ax_Ay_50_28_140}}
{Ax_Ay_45_24_for_Paper}
\hvFloat[
 floatPos=!htp,
 capWidth=h,
 capPos=r,
 capAngle=90,
 objectAngle=90,
 capVPos=c,
 objectPos=c]
{figure*}{\includegraphics[width=1.4\textwidth]{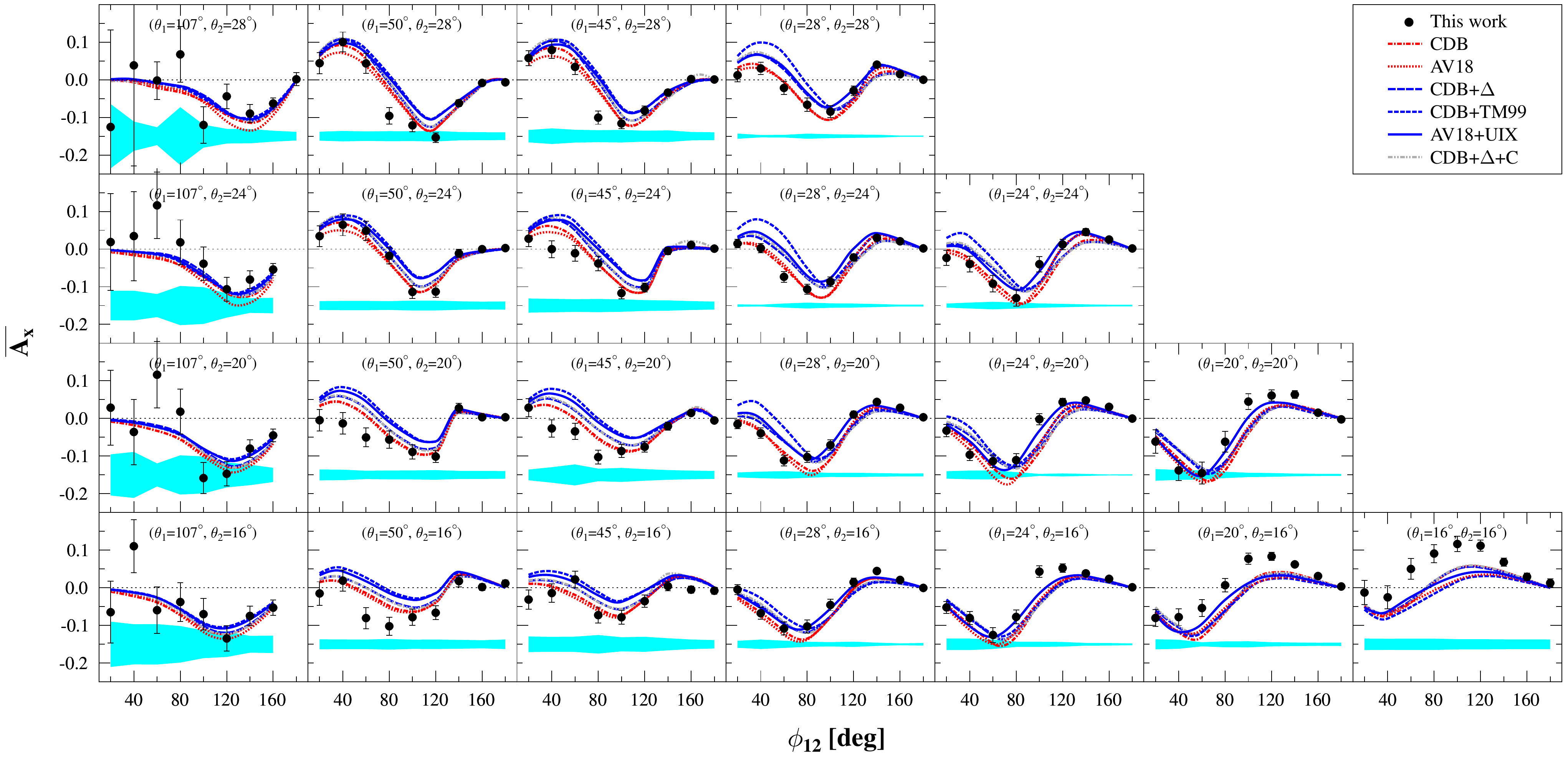}}
{The $A_{x}$ analyzing powers, averaged over $S$ for every kinematics configuration 
$(\theta_{1},\theta_{2})$ versus $\phi_{12}$ for backward-forward and forward-forward configurations.
The experimental data of forward-forward configurations $(\theta_{1}\leq 28^{\circ},\theta_{2}\leq 28^{\circ})$
are taken from Ref.~\cite{Hajar_thesis}. Similarly, the average analyzing powers for 
the theoretical predictions are produced and are presented by the same line styles 
as were chosen for the lines depicted in Fig.~\ref{Ax_Ay_50_28_140}. 
The cyan bands depict $2\sigma$ systematic uncertainties.}
{Ax_weighted_average_bw_wwHajar}
\hvFloat[
 floatPos=!htp,
 capWidth=h,
 capPos=r,
 capAngle=90,
 objectAngle=90,
 capVPos=c,
 objectPos=c]
{figure*}{\includegraphics[width=1.4\textwidth]{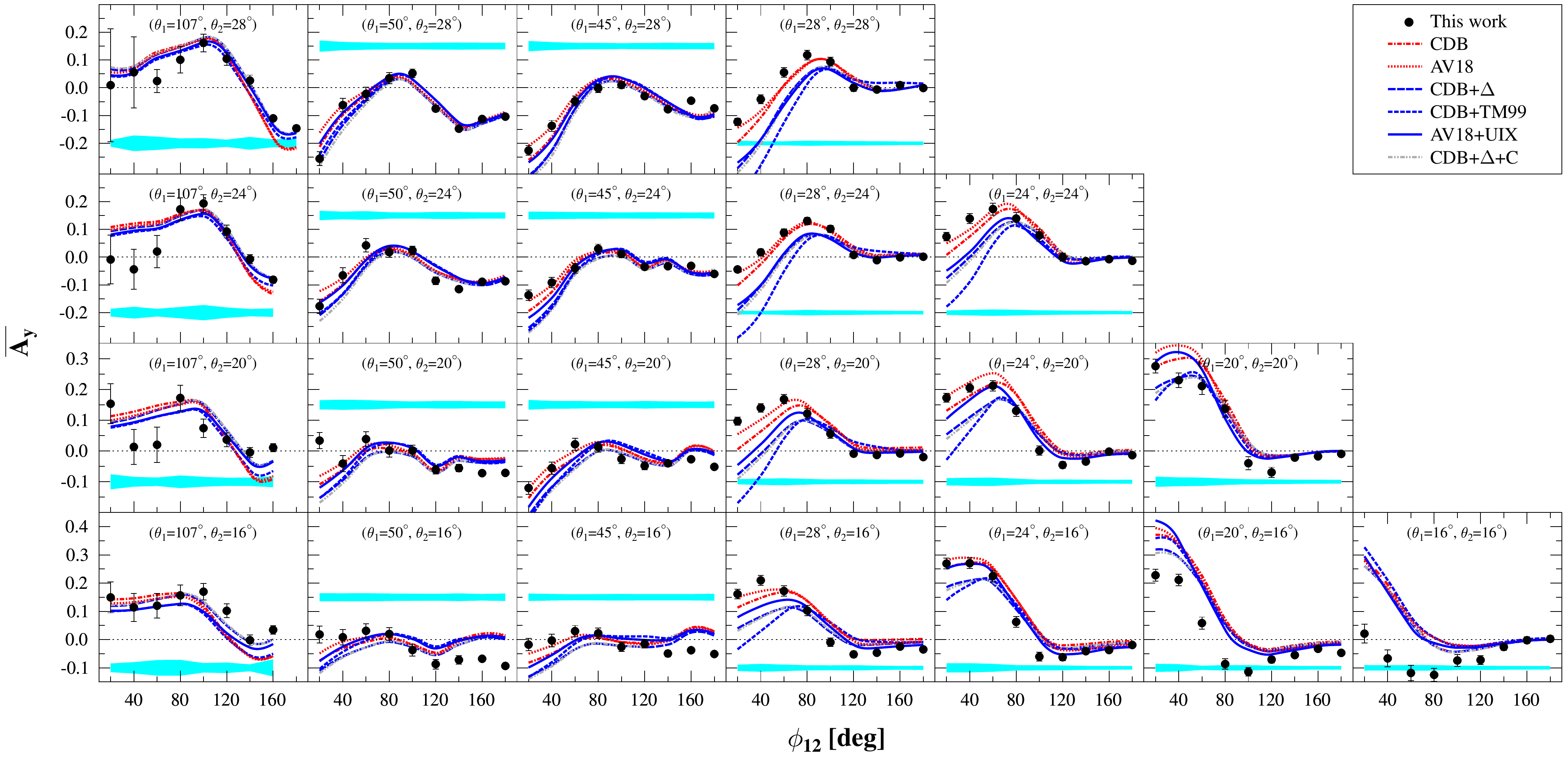}}
{Same as Fig.~\ref{Ax_weighted_average_bw_wwHajar} except for $A_{y}$.}
{Ay_weighted_average_bw_wwHajar}
\section{Discussion}
\label{Dis}
To improve our insight into 3NF effects and to monitor the consistency of the results, we decided to do a systematic survey of all experimental analyzing powers which are obtained up to now for the proton-deuteron break-up reaction at $135$~MeV with BINA. 

For the survey, we studied the overall progression of the measured analyzing powers for forward-forward and forward-backward configurations. Figures~\ref{Ax_weighted_average_bw_wwHajar} and~\ref{Ay_weighted_average_bw_wwHajar} depict this progression for the analyzing powers averaged over $S$ as a function of $\phi_{12}$. In addition, the predictions by theoretical calculations based on a variety of input potentials are presented by lines. 

In general, we observe that the state-of-the art calculations describe the data well for a large part of the phase space. However, significant discrepancies between data and theory can be observed in particular for $A_y$ at small $\phi_{12}$ corresponding to small relative energies between the two final-state protons. These discrepancies cannot be explained by the Coulomb effect, neither by the 3NF effect originating from the $\Delta$ resonance.

From a more detailed inspection of the $A_{x}$ results presented in Fig.~\ref{Ax_weighted_average_bw_wwHajar}, we can observe the following aspects:
\begin{enumerate}
\item Towards $\phi_{12}=180^{\circ}$, $A_{x}$ is measured to be zero as expected under parity conservation. This is indeed compatible with our data lending confidence to our procedure to extract this observable.
\item For data taken at $\theta_{1}=107^{\circ}$, the statistical and systematical uncertainties are very large and hence the data are not sensitive to studying the details of the 3N interaction. For all other configurations our data show sensitivity (errors are smaller than model deviations).
\item The model sensitivity is the largest at configurations that are away from coplanarity. In general, it appears that calculations that incorporate the 3NF effects result in a worse description of the data, albeit small in most cases.
\end{enumerate}

By reviewing the $A_{y}$ results depicted in Fig.~\ref{Ay_weighted_average_bw_wwHajar}, we draw the following conclusions:
\begin{enumerate}
\item The model sensitivity is significantly larger in $A_{y}$ than in $A_{x}$, in particular towards non-coplanarity and for moderate scattering angles of the two protons.
\item The failure of the models that incorporate 3NF is very evident for this observable. Strikingly, the calculations are more compatible with data when no 3NF is taken into account. A similar problem was observed in an experiment with BINA at a beam energy of 190 MeV~\cite{Mardanpour10,Maisam_2020}. Hence, the current 3NF models appear to miss an important ingredient to describe this observable at various intermediate energies below the pion-production threshold.
\item For symmetric configurations, $\theta_{1}=\theta_{2}$ and at large $\phi_{12}$, the average of $A_{y}$ should become zero because of symmetry arguments. This is confirmed by the data. The problem with description of this observable by currently available models is also evident.
\end{enumerate}

A discrepancy\textcolor{referee1}{, similar to the one observed in Refs.~\cite{Mardanpour10,Maisam_2020},} between the measured analyzing powers and theoretical predictions arises for close-to-symmetric configurations at small scattering angles of the two final-state protons. These particular cases were studied in more detail in the past and a discussion can be found in Refs.~\cite{Mardanpour10,Hajar_2020}. It has been speculated that for these configurations, the two protons are in a relative S wave, corresponding to the $d(\pol{p},^2$He$)n$ reaction. By comparing the results of the $d(\pol{p},^2$He$)n$ channel with $d(\pol{p},p)d$ scattering, one might conclude that the discrepancy is related to a spin-isospin deficiency of the 3NF models.
\section{Summary and conclusion}
\label{summary}
Our measurements cover a large part of the total phase space of the break-up reaction. This allowed us to study systematically the two vector analyzing powers, $A_{x}$ and $A_{y}$, for various scattering angles and with respect to the full range of coplanarity of the two final-state protons. The data were compared to state-of-art Faddeev calculations that were based on several NN and 3NF models. With such a large coverage, we were able to significantly expand the previously-published and experimentally-probed phase space. Moreover, with our measurements we were able to probe parts of the break-up phase space at which one expects to have no sensitivity to 3NF effects and parts at which the predictions significantly vary depending on the choice of input potential. In general, we observed that the calculations are compatible with the data at configurations with low-model sensitivity. Strikingly, though, at places with a strong model sensitivity, the calculations that include a 3NF effect give a significantly worse description of the data compared to the results that excludes a 3NF effect. The model deficiency appears to be the strongest for the observable $A_{y}$, giving rise to another $A_{y}$ puzzle in the proton-deuteron break-up channel at intermediate energies. The origin of the observed discrepancy is yet unknown and requires a further theoretical study.
\section*{Acknowledgement}
\label{Ack}
The authors acknowledge the work by AGOR cyclotron and ion-source groups at KVI for delivering the high-quality polarized beam. This work was partly supported by the Polish National Science Centre under Grant No. 2012/05/B/ST2/02556 and 2016/22/M/ST2/00173. The numerical calculations were partially performed on the supercomputer cluster of the JSC, J\"ulich, Germany.


\bibliographystyle{my_epj}
\bibliography{References}

\end{document}